\documentclass[twocolumn]{aastex62}

\usepackage{amsmath}
\usepackage{blindtext}
\usepackage{color}

\def\aap{{A\&A}}
\def\aj{{AJ}}
\def\apj{{ApJ}}
\def\apjl{{ApJ}}
\def\araa{{ARA\&A}}
\def\mnras{{MNRAS}}
\def\nat {{Nature}}

\def\apjs {{ApJS}}

\def\pasj {{Publications of the Astronomical Society of Japan}}

\def\Msun {\,\mathrm{M}_\odot}
\def\ccm {\,\mathrm{cm}^{-3}}
\def\tcoll {t_\mathrm{coll}}

\shorttitle{CEMP Stars as Consequence of Inhomogeneous Mixing}
\shortauthors{T. Hartwig \& N. Yoshida}

\begin{document}

\title{Formation of carbon-enhanced metal-poor stars as a consequence of inhomogeneous metal mixing}

\correspondingauthor{Tilman Hartwig}
\email{Tilman.Hartwig@ipmu.jp}

\author{Tilman Hartwig}
\affil{Kavli IPMU (WPI), UTIAS, The University of Tokyo, Kashiwa, Chiba 277-8583, Japan}
\affil{Department of Physics, School of Science, The University of Tokyo, Bunkyo, Tokyo 113-0033, Japan}
\affil{Institute for Physics of Intelligence, School of Science, The University of Tokyo, Bunkyo, Tokyo 113-0033, Japan}
\author{Naoki Yoshida}
\affil{Kavli IPMU (WPI), UTIAS, The University of Tokyo, Kashiwa, Chiba 277-8583, Japan}
\affil{Department of Physics, School of Science, The University of Tokyo, Bunkyo, Tokyo 113-0033, Japan}
\affil{Institute for Physics of Intelligence, School of Science, The University of Tokyo, Bunkyo, Tokyo 113-0033, Japan}

\begin{abstract}
We present a novel scenario for the formation of carbon-enhanced metal-poor (CEMP) stars. Carbon enhancement at low stellar metallicities is usually considered a consequence of faint or other exotic supernovae. An analytical estimate of cooling times in low-metallicity gas demonstrates a natural bias, which favours the formation of CEMP stars as a consequence of inhomogeneous metal mixing: carbon-rich gas has a shorter cooling time and can form stars prior to a potential nearby pocket of carbon-normal gas, in which star formation is then suppressed due to energetic photons from the carbon-enhanced protostars. We demonstrate that this scenario provides a natural formation mechanism for CEMP stars from carbon-normal supernovae, if inhomogeneous metal mixing provides carbonicity differences of at least one order of magnitude separated by $>10$\,pc. In our fiducial (optimistic) model, $8\%$ ($83\%$) of observed CEMP-no stars ([Ba/Fe] $<0$) can be explained by this formation channel. This new scenario may change our understanding of the first supernovae and thereby our concept of the first stars. Future 3D simulations are required to assess the likelihood of this mechanism to occur in typical high-redshift galaxies.
\end{abstract}

\keywords{early Universe --- stars: Population II --- stars: Population III --- Local Group --- stars: abundances}

\section{Introduction} \label{sec:intro}
The first stars in the Universe formed out of primordial gas when the Universe was only a few hundred million years old. The characteristic mass of these so-called Population III (Pop~III) stars is higher than for present-day star formation due to the lack of metals, which otherwise provide efficient cooling channels \citep{bromm99,omukai05}.

The chemical fingerprint of the first stars is preserved in the metal abundances of second-generation stars: high-resolution spectroscopy of such extremely metal-poor (EMP) stars in the Milky Way allows us to determine the properties of their formation environment and progenitor stars \citep{frebel15,ishigaki18,hartwig18c}.

A special feature of EMP stars is that [C/Fe]\footnote{Defined as $[\mathrm{A}/\mathrm{B}] = \log_{10}(m_\mathrm{A}/m_\mathrm{B})-\log_{10}(m_{\mathrm{A},\odot}/m_{\mathrm{B},\odot})$, where $m_\mathrm{A}$ and $m_\mathrm{B}$, are the abundances of elements A and B and $m_{\mathrm{A},\odot}$ and $m_{\mathrm{B},\odot}$ are the solar abundances of these elements \citep{asplund09}.} increases with decreasing metallicity \citep{beers92,aoki07,lee13,sharma16} with all stars having [C/Fe] $>0.7$ at [Fe/H] $\lesssim -4$ \citep{yong13,placco14}. Empirically, there are different subclasses of so-called carbon-enhanced metal-poor (CEMP) stars ([C/Fe] $>0.7$), mainly classified by their abundance of neutron capture elements \citep{beers05}.
Here, we focus on CEMP-no stars ([Ba/Fe] $<0$), which are potentially enriched by SNe \citep{tominaga07} or by mass transfer in a stellar binary \citep{arentsen18}.

Carbon enhancement is specific for EMP stars at very low metallicities and therefore directly related to the properties of SNe and star formation in the early Universe. Whereas CEMP-s stars may have accreted additional elements during their stellar lifetimes \citep{lucatello05}, CEMP-no stars are expected to reflect the chemical composition of their birth environment. Conventional Pop~III core-collapse SNe and high-mass pair-instability SNe produce [C/Fe] $<0.7$ \citep{nomoto13}. One possible explanation for CEMP-no stars at low metallicities are faint SNe \citep{iwamoto05,tominaga07,ishigaki14}. These SNe eject less iron, which results in an increased [C/Fe]. If about half of the Pop~III core-collapse SNe are faint, theoretical models can reproduce the fraction of CEMP stars at low metallicities \citep{ji15,deB17,hartwig18}.

It is conventionally assumed that the observed elemental ratios of EMP stars correspond to the elemental ratios from the enriching SN. However, what if the observed [C/Fe] value does not correspond to the ratio produced by the progenitor SN? Already small amounts of inhomogeneous metal mixing can lead to an observable bias, specifically toward the preferred formation of CEMP stars at low metallicities.

Little research has been conducted on the possible effects of inhomogeneous mixing of different elements. For a core-collapse SN, Rayleigh Taylor instabilities develop and mix different elements during the homologous expansion of the SN \citep{joggerst10}. Simulations of metal enrichment in the first galaxies also suggest that metals are well mixed with the interstellar medium after fall-back of the SN explosion \citep{greif07,jeon14,chiaki18}.

In contrast, 3D simulations that follow different representative SN shells with Lagrangian tracer particles find that different element groups are not homogeneously mixed after recollapse of the gas \citep{ritter15,sluder16}. \citet{joggerst11} demonstrate with 2D models that pair-instability SNe have well defined shells of different elements prior to explosion. They find no substantial mixing of these elements to times well after shock breakout from the surface of the star. Moreover, asymmetric or jet-induced SNe can eject metals into different directions with element-specific efficiencies \citep{maeda02,tominaga09,wongwathanarat13}. \citet{sluder16} find elemental abundance ratios in the recollapsing gas (e.g. [C/Fe]) that differ by about one order of magnitude from the original SN. They argue that such inhomogeneous mixing could be responsible for the large diversity and scatter of elemental abundances at low metallicities.

\section{Methodology}
\begin{figure*}
\plotone{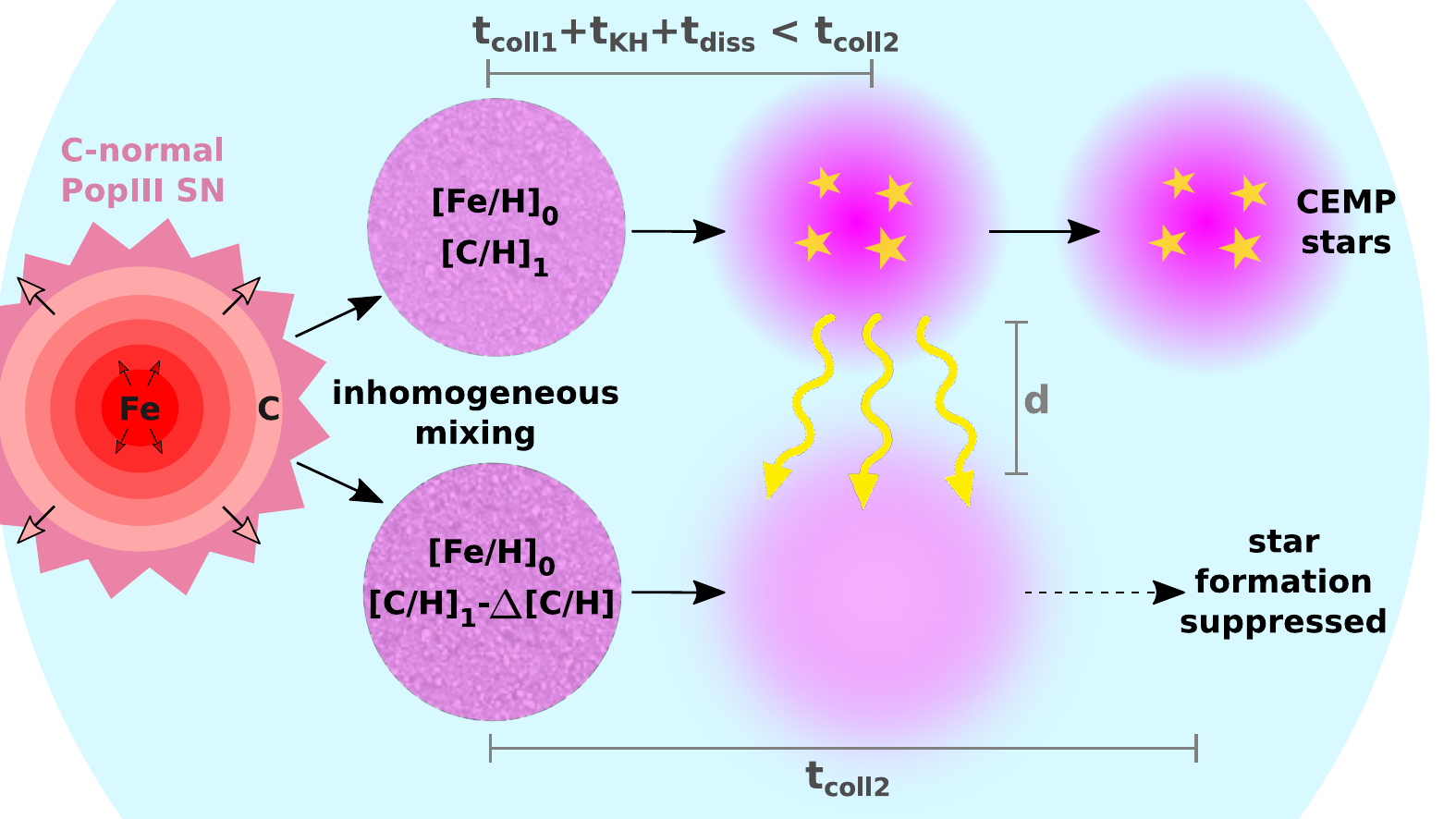}
\caption{Illustration of how a C-normal SN ([C/Fe] $<0.7$) can trigger the formation of CEMP stars ([C/Fe] $>0.7$). The crucial quantities are the separation of the two clumps, $d$, and their carbonicity difference, $\Delta$[C/H].\label{fig:illu}}
\end{figure*}

The main idea of the proposed scenario is illustrated in Figure~\ref{fig:illu}.
After a Pop~III SN, gas recollapses and triggers the formation of second-generation stars \citep{chiaki18}. During the recollapse, multiple overdensities can form (M. Magg et al. 2019, in preparation) with potentially different elemental abundances (top and bottom row in Figure~\ref{fig:illu}). If the collapse time, $t_\mathrm{coll}$, of the carbon-rich clump is short enough, this clump can form stars and prevent star formation in the nearby clump by its energetic photons. We derive the analytical condition for this scenario in the following sections.\footnote{The corresponding Mathematica notebook to reproduce the results can be found here: \url{https://gitlab.com/thartwig/CEMP_InhomogeneousMixing}}

The initial condition of two separated clumps with different chemical compositions can also be realised in other ways: \citet{pan12} demonstrate how SN ejecta can mix into existing molecular gas, at the periphery of an \ion{H}{2} region. However, the hard radiation from the first stars is expected to evacuate most gas from their host halo \citep{schauer17}, which limits the feasibility of this specific constellation. Another way is external enrichment from a nearby, star-forming halo. We expect such external enrichment to occur in only $\lesssim 10\%$ of second-generation star formation events \citep{hartwig18} and to be inhomogeneous \citep{sarmento17}. Although the ejected elements have longer time to mix after an external enrichment, they are homogeneously mixed only on subparsec scales \citep{smith15}.

\subsection{Cooling}
To understand the relevant time scales, we first summarize the cooling properties of metal-poor gas.
The thermal evolution of metal-poor gas starts to deviate from the metal-free thermal evolution at densities of around $\sim 10^3\ccm$ \citep{schneider12a}. At these metallicities of $\sim 10^{-3}$ solar, dust only has a significant influence at $n \gtrsim 10^5 \ccm$ \citep{omukai05}. For lower metallicities, dust becomes only relevant at $\gtrsim 10^{10}\ccm$ \citep{schneider12a}. Also the formation of molecular hydrogen, H$_2$, on dust grains becomes only important at metallicities of $Z>10^{-4}$ solar \citep{omukai00}. At these metallicities, H$_2$ cooling plays only a subordinate role.


\ion{C}{2} and \ion{O}{1} line cooling are the dominant cooling channels in gas with $Z \approx 10^{-2}$ up to densities of $\sim10^4\ccm$ in halos with a moderate UV background \citep{bromm03,schneider12a}. We only consider \ion{C}{2} cooling here, because it is more efficient in our fiducial model \citep{sp05} and because carbon is easier to observe in the atmospheres of EMP stars.
We multiply the cooling rate at solar carbonicity \citep{sp05}
\begin{equation}
\frac{\Lambda _{\mathrm{CII}}}{\mathrm{erg}\,\mathrm{cm}^{-3}\,\mathrm{s}^{-1}} = 4.8 \times 10^{-21} \left( \frac{n}{10^3\ccm} \right)^2 \exp \left( -\frac{92\,\mathrm{K}}{T} \right)
\end{equation}
with $10^{[\mathrm{C/H}]}$ to obtain a carbonicity-dependant cooling rate, which is valid for densities of $\lesssim 3000\ccm$. We also include cooling by H$_2$, with a fiducial abundance of $f_\mathrm{H2}=10^{-3}$, which becomes the dominant cooling channel at very low carbonicities \citep{gallipalla98}.

Primordial gas reaches a so-called loitering state at $n = 10^3-10^4\,\mathrm{cm}^{-3}$ with typical temperatures of $\sim 200$\,K, before it proceeds with runaway contraction \citep{omukai05}. Following \citet{frebel07}, we also choose this state with $n = 10^3\,\mathrm{cm}^{-3}$ and $T=200$\,K in metal-poor gas as initial conditions for the two clumps before the carbon-induced collapse starts to take over. Under these conditions, the collapse of a second clump could still be reversed by a nearby ($\lesssim 100$\,pc) photodissociating UV source \citep{susa07}. We demonstrate below how the choice of the initial conditions affects the results.




To focus on the bias that is introduced by a different carbonicity as a consequence of inhomogeneous metal mixing, we only consider clumps that have the same density. In reality, one cloud could be more or less dense than the other. If we assume that [C/Fe] is independent of the gas density, two equal-density clumps can be seen as the average scenario, marginalised over cloud pairs with different initial densities.



\subsection{Kelvin-Helmholtz (KH) Time}
A newly formed protostar reaches the main sequence on the KH time, where it starts to produce energetic photons. The KH time decreases with stellar mass and more massive stars are therefore the first objects to ignite hydrogen burning. The carbon-rich clump will form a cluster of stars with various masses. We assume that the star that first forms in the carbon-rich gas has $25\Msun$ and discuss the influence of this choice on the photodissociation time below. Following \citet{sp05}, the KH time is given by
\begin{equation}
t_\mathrm{KH}=30\,\mathrm{Myr} \left( \frac{M_*}{\Msun} \right)^{2} \left( \frac{R_*}{\mathrm{R_\odot}} \right)^{-1} \left( \frac{L_*}{\mathrm{L}_\odot} \right)^{-1},
\end{equation}
with stellar mass $M_*$, radius $R_*$, and luminosity $L_*$. For a $25\Msun$ star, the KH time is $1.3\times 10^4$\,yr, which is short compared to other involved time scales.

\subsection{Photodissociation Time}
Massive stars produce a strong Lyman--Werner (LW) flux that can photodissociate H$_2$, the main coolant of metal-poor gas at $n \lesssim 10^3\ccm$.


\citet{glover01} derive the H$_2$ photodissociation time for a dense clump by a nearby star. A $25\Msun$ star provides an LW photon emissivity of $\dot{N}_\mathrm{LW}=10^{48}\,\mathrm{s}^{-1}$ \citep{bromm01,sp05}. Assuming a spherical clump of mass  $M_\mathrm{clump}$ and density $n$, in a distance of $d$ from the star, an H$_2$ fraction of $f_\mathrm{H2}=10^{-3}$, an average dissociation probability per LW photon of $f_\mathrm{dis}=0.2$, and the fraction of LW photons that are absorbed by the clump, $f_\mathrm{abs} = 0.05$, \citet{glover01} derive the photodissociation time
\begin{equation}
t_\mathrm{dis}=0.2\,\mathrm{Myr}\,n_3^{2/3} d_{10}^2 \left( \frac{M_\mathrm{clump}}{1000\Msun} \right)^{1/3},
\end{equation}
where $n_3=n/1000\ccm$ and $d_{10}=d/10$\,pc.
%
The strongest dependence is on the separation of the second cloud from the newly formed star. The typical cloud mass of $\sim 1000\Msun$ corresponds to the Jeans mass at our initial conditions. The minimum separation of two distinct molecular clouds of $\sim 1000\Msun$ is twice their typical radius of $\sim 10$\,pc. For the maximum separation of two molecular clouds, we assume $\sim 100$\,pc, which is of the order of the virial radius of high-redshift minihalos. We therefore choose an intermediate separation of $d=30$\,pc for our fiducial model.

The photodissociation of H$_2$ and additional photoheating by dust in the second clump will delay its collapse \citep{nakatani18}. Ionization of the second clump will then revert the collapse and eventually prevent star formation.


\subsection{Cooling Condition}
The condition that a carbon-rich cloud (1) can collapse first and the newly formed star suppress star formation in a nearby, carbon-normal clump (2) is
\begin{equation}
t_\mathrm{coll,1}+t_\mathrm{KH}+t_\mathrm{diss} < t_\mathrm{coll,2}
\label{eq:condition}
\end{equation}
with the carbonicity dependent collapse times $t_\mathrm{coll,1}=t_\mathrm{coll,1}([\mathrm{C/H}])$ and $t_\mathrm{coll,2}=t_\mathrm{coll,2}([\mathrm{C/H}]-\Delta [\mathrm{C/H}])$. Phrased differently, cloud 1 has to dissociate the H$_2$ in cloud 2 before the latter one has collapsed. We refer to the carbonicity of clump 1 as [C/H] and assume that the metallicity of the second clump is lower by $\Delta [\mathrm{C/H}]$.
We estimate the collapse time as
\begin{equation}
\tcoll=\mathrm{max}(t_\mathrm{ff},t_\mathrm{cool})
\end{equation}
with the freefall time $t_\mathrm{ff}$ and the cooling time
\begin{equation}
t_\mathrm{cool} = \frac{u}{\Lambda_\mathrm{CII}+\Lambda_\mathrm{H2}},
\end{equation}
with the internal energy $u = k_B T n/(\gamma -1)$. This provides similar results to the one-zone model by \citet{omukai05} in the limit of low metallicity ($\tcoll=t_\mathrm{cool}$) and high metallicity ($\tcoll=t_\mathrm{ff}$).

\begin{figure}
\plotone{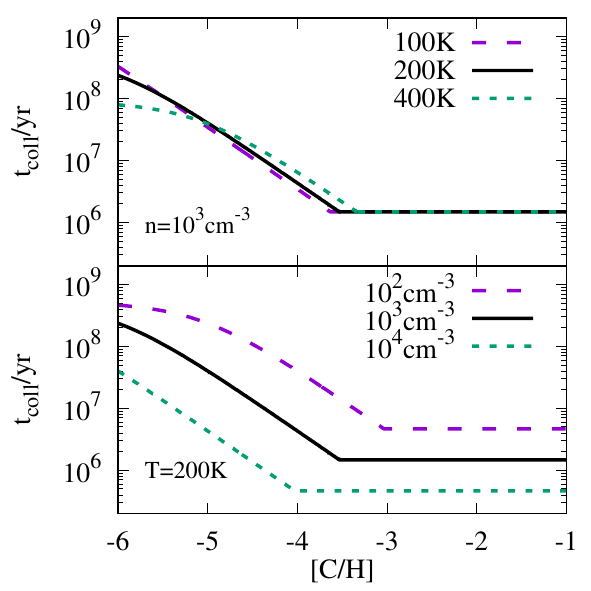}
\caption{Dependency of the collapse time on the carbonicity for different values of the temperature (top) and density (bottom). The collapse time depends more strongly on the initial density than on the initial temperature.\label{fig:3CoH}}
\end{figure}

With this closed set of equations, we calculate $\tcoll$ as a function of the carbonicity, which is illustrated for different values of the initial density and temperature in Figure~\ref{fig:3CoH}.
Below [C/H]$_\mathrm{crit} \lesssim -3.6$ the cooling is inefficient and the collapse time is limited by the cooling time. At higher carbonicity, the collapse proceeds isothermally and the collapse time approaches the freefall time at the corresponding density. We assume that the characteristic mass of second-generation stars is small enough that most of them have survived until today. If, however, all EMP stars below a certain metallicity are more massive than $\sim 0.8\Msun$ they do not survive until today and no additional mechanism, such as photodissociation, is required to prevent their existence in the present-day Milky Way.

For the proposed scenario to work, the carbon-normal clump must have a carbonicity below [C/H]$_\mathrm{crit}$ and observationally most CEMP-no stars are above [C/H]$_\mathrm{crit}$. In the most common case, where the carbon-rich clump is above and the carbon-normal clump below [C/H]$_\mathrm{crit}$, we can provide a closed analytical expression for the required inhomogeneity:
\begin{equation}
\begin{aligned}
&\Delta [\mathrm{C/H}] > [\mathrm{C/H}] + 3.6 + \\
&\log _{10}\left( \frac{13000+2 \times 10^5 d_{10}^2 n_3^{2/3}+1.5\times 10^6 n_3 ^{-1/2}}{1.7 \times 10^6} n_3 \right),
\end{aligned}
\end{equation}
where we have assumed $T=200$\,K, $M_*=25\Msun$, and that \ion{C}{2} cooling dominates.



\section{Results}
\begin{figure}
\centering
\plotone{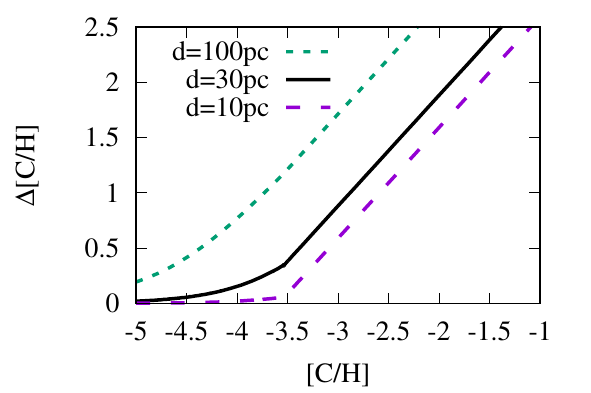}
\caption{Required inhomogeniety as a function of the carbonicity of the potential CEMP star for different clump separations. For $d=30$\,pc and a carbonicity of [C/H] $\approx -4$, an inhomogeniety as small as $\sim 0.2$\,dex is sufficient to trigger this formation scenario. At higher carbonicities or larger separations, more inhomogeneous mixing of metals is required.\label{fig:CoHdCoH}}
\end{figure}

We show the necessary inhomogeniety as a function of the carbonicity in Figure~\ref{fig:CoHdCoH}.
The KH and photodissociation time are independent of the carbonicity. The collapse time becomes shorter with decreasing carbonicity due to less efficient cooling. Therefore, already a small difference in the carbonicity of $\Delta[\mathrm{C/H}] \approx 0.2$\,dex at [C/H] $\leq -4$ is sufficient to compensate for the KH and photodissociation time. If the clumps are separated further, the photodissociation time increases and therefore also the required inhomogeniety increases. If we find a CEMP star above the illustrated lines, it could have formed from a C-normal SN.

\begin{figure}
\centering
\plotone{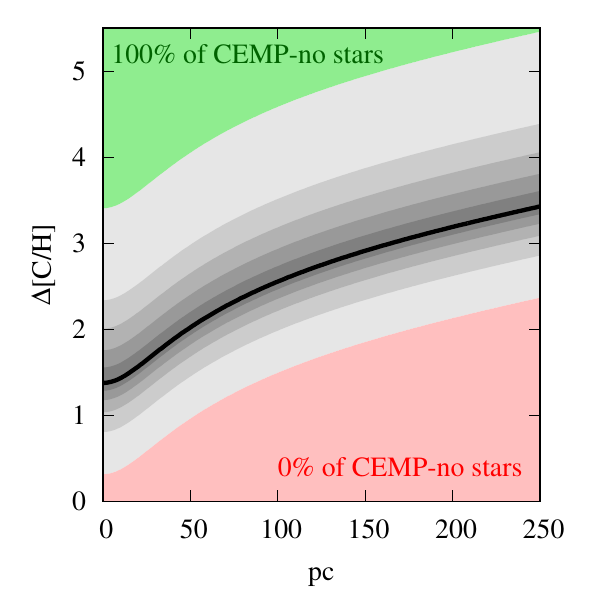}
\caption{Every observed CEMP-no star could be represented by one line in this plot that yields the minimum required carbonicity difference, $\Delta$[C/H], of two clumps separated by a certain distance for this proposed scenario to work. The plot illustrates the percentiles for all observed CEMP-no stars in grey. If all molecular cloud pairs in the first galaxies would be in the green area, 100\% of the observed CEMP-no stars could result from C-normal SNe as a consequence of inhomogeneous metal mixing. If all pairs would be in the red area, metal mixing would be too homogeneous to explain any CEMP-no star by this scenario.\label{fig:percentiles}}
\end{figure}

We provide a direct comparison with observations and estimate the likelihood of this scenario to occur in Figure~\ref{fig:percentiles}.
To determine which level of inhomogeneity we can expect on which spatial scale, requires improved 3D simulations that follow the mixing of different elements after an SN explosion. This figure turns the question around and predicts which fraction of the currently observed CEMP-no stars could be explained by inhomogeneous metal mixing after a C-normal SN. For example, if two clumps with a separation of $\sim 50$\,pc have $\Delta$[C/H] $\approx 2$, 50\% of observed CEMP-no stars can be explained with this scenario (black line). The other 50\% would require higher inhomogeneity. This comparison to observations is based on 141 CEMP-no stars with [C/Fe] $>0.7$ and [Ba/Fe] $<0$ (J. Yoon, priv. communication).

\begin{figure}
\centering
\plotone{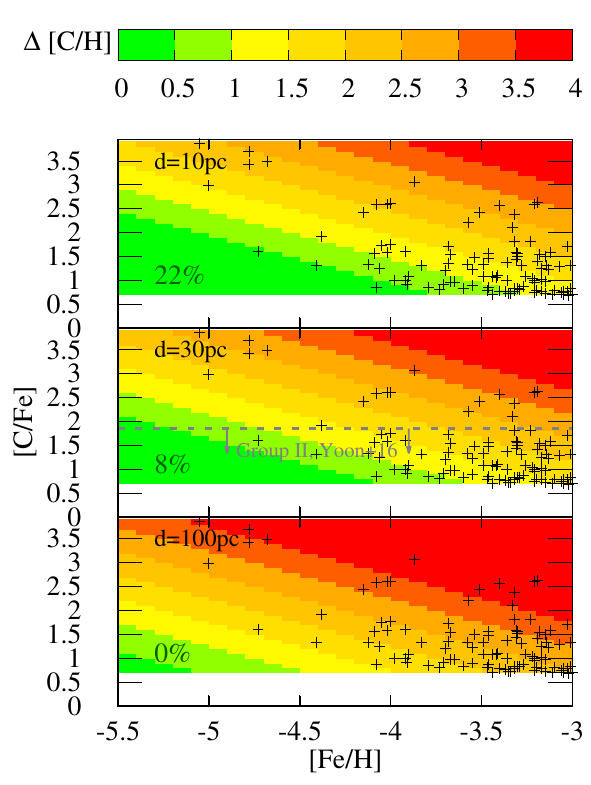}
\caption{The color indicates the minimum carbonicity difference required for a CEMP-no star to form via the proposed channel from a C-normal SN. The three panels illustrate the results for typical separations of star-forming clumps of 10\,pc (top), 30\,pc (middle), and 100\,pc (bottom) and the black symbols represent observed CEMP-no stars, based on \citet{abohalima17}. The percentage in each panel corresponds to the fraction of observed CEMP-no stars that could be explained by this scenario under the assumption of $\Delta$[C/H]$\approx 1$\,dex. In the fiducial model, all CEMP-no stars that could form via the proposed channel are in `Group II' according to the classification by \citet{yoon16} because `Group III' stars require even higher inhomogenieties.\label{fig:sum3}}
\end{figure}

We compare our prediction to individual observed CEMP-no stars in Figure~\ref{fig:sum3}.
In our fiducial model, where clumps are separated on average by $30$\,pc with inhomogeneities of the order $\Delta$[C/H] $\approx 1$\,dex, we can explain $8\%$ of the currently observed CEMP-no stars via this formation channel (green). If clouds are separated by $100$\,pc, $\Delta$[C/H] $\approx 1$\,dex could account for no observed stars and if clouds are separated by $10$\,pc, $22\%$ of observations could be explained. If we find $\Delta$[C/H] $\approx 2$\,dex at typical separations of $10$\,pc, up to $83\%$ of CEMP-no stars can form via this channel.



\section{Discussion and Conclusion}
We propose a novel formation scenario for CEMP stars as a consequence of inhomogeneous metal mixing after the first SNe. C-rich gas has a shorter cooling and therefore collapse time than C-normal gas. We demonstrate analytically that there is a bias favouring the formation of CEMP stars and the supression of C-normal stars if two clumps form with different carbonicities (and the same iron content). The two crucial parameters are the distance of these two clumps and their difference in carbonicity. In our fiducial model with $\Delta$[C/H]$\approx 1$\,dex and a clump separation of $30$\,pc we find that $8\%$ of observed CEMP-no stars could have formed via this proposed pathway. For closer clumps and therefore shorter photodissociation times, up to $22\%$ of CEMP stars could result from this scenario.

We do not claim that all CEMP-no stars have formed as a consequence of inhomogenious metal mixing from C-normal SNe. However, a certain fraction of CEMP-no stars may have formed in this way, which limits the fraction of faint SNe in the early Universe, which current models predict to be $\gtrsim 50\%$ \citep{ji15,deB17} or $\sim 10\%$ \citep{ishigaki18}.

The proposed bias is a natural consequence of the cooling properties of metal-poor gas. This insight provides another valuable observational consequence: if inhomogeneous mixing occurs and only the clumps with higher cooling rates succeed to form low-mass stars that survive until today, we should see a similar trend not only for C-rich stars, but also for other elements that are efficient coolants in metal-poor gas. Indeed, there is a similar trend for oxygen \citep{saga,abohalima17}, which provides cooling via its atomic (\ion{O}{1}, or molecular (OH, CO) line transitions, or as constituents of silicate dust at higher densities \citep{schneider12a,chiaki17}. The inclusion of oxygen cooling will shift the required inhomogeneities to slightly lower overall metallicities, but qualitatively will not change the results as it affects the cooling properties in both clumps.

We assume that the number densities of iron and hydrogen of the two clumps are the same and that the number densities of carbon are different. To produce CEMP stars, inhomogeneous mixing needs to increase [C/Fe] from the SN explosion to star formation. To introduce a bias via different cooling times, the carbonicity [C/H] needs to be higher in one clump. These two conditions could also be reached with homogeneous mixing of carbon, inhomogeneous mixing of iron, and different hydrogen densities in the two clumps.

We assume that the star that forms first in the recollapsed gas has $25\Msun$. If instead a $10\Msun$ star forms, it has a longer KH time of $t_\mathrm{KH}=0.1$\,Myr and its LW emissivity is one order of magnitude smaller than that of a $25\Msun$ star, with a correspondingly longer photodissociation time. In this case, only 5\% (0\%) of the currently observed CEMP-no stars could be explained with the proposed scenario for typical clump separations of 10\,pc (30\,pc). If we assume instead a higher initial density of $10^4\ccm$, the collapse time is shorter and the carbonicity difference needs to be larger to account for this effect. Consequently, CEMP-no stars could be explained only if $\Delta$[C/H]$\gtrsim 2$\,dex for a typical clump separation of 30\,pc.

Next-generation 3D simulations will provide estimates on the mixing efficiency of different elements after a Pop~III SN. This information is not only crucial as the basis for this scenario, but more generally important to answer the question of to what extent observed chemical abundance ratios of an EMP star correspond to the chemical abundances of the enriching SN.

\acknowledgments
TH is a JSPS International Research Fellow. We thank G. Chiaki, J. Yoon, R. Nakatani, and K. Chen for stimulating discussions. We appreciate valuable comments by the anonymous referee and Simon Glover on this manuscript. We are deeply thankful to Jinmi Yoon for sharing her compilation of observed CEMP-no stars with us. This work was supported by World Premier International Research Center Initiative (WPI Initiative), MEXT, Japan.

\end{document}